\documentclass[prl,twocolumn,tightenlines]{revtex4-1}
\usepackage{tipa}
\usepackage{mathrsfs}
\usepackage{amsmath}
\usepackage{amssymb}
\usepackage{graphicx}

\newcommand{\be}{\begin{eqnarray}}
\newcommand{\ee}{\end{eqnarray}}

 \newcommand{\gsim}{\mathrel{\hbox{\rlap{\lower.55ex \hbox {$\sim$}}
                   \kern-.3em \raise.4ex \hbox{$>$}}}}
\newcommand{\lsim}{\mathrel{\hbox{\rlap{\lower.55ex \hbox {$\sim$}}
                   \kern-.3em \raise.4ex \hbox{$<$}}}}

\begin{document}

\title{ Studies of the hydrodynamic evolution of the dense baryonic matter produced in relativistic heavy ion collisions }

\author{Gong-Ming Yu}
\email{ygmanan@163.com}

\author{Jian-Song Wang}
\email{jswang@impcas.ac.cn}

\address{ Institute of Modern Physics, Chinese Academy of Sciences, Lanzhou 730000, China }

\begin{abstract}
We use the relativistic perfect-fluid hydrodynamics to describe the space-time evolution of dense baryonic matter produced in the central nucleus-nucleus collisions at HIRFL-CSR, HIAF, FAIR-CBM, NICA-MPD, and RHIC-BES. The transverse flow of the fireball with cylindrical symmetry and boost invariant along the longitudinal direction is also analyzed. Based on the relativistic kinetic theory, we also present the first preliminary calculation on the production of low-mass dileptons and low-$p_{T}$ photons can be considered as the probes of the dense baryonic matter produced in the central Au-Au and U-U collisions. We find that the rapid cooling of the expanding dense baryonic matter with transverse flow effect can lead to the suppression of the low-mass dileptons and low-$p_{T}$ photons production at HIRFL-CSR, HIAF, FAIR-CBM, NICA-MPD, and RHIC-BES energies.


\end{abstract}

\maketitle

High energy heavy-ion collision experiments at relativistic energies can create extreme states of strongly interacting matter and enable their investigation in the laboratory. Experiments at Large Hadron Collider (LHC) and top Relativistic Heavy Ion Collider (RHIC) energies can explore the quantum chromodynamics (QCD) phase diagram in the transition region between quark matter and hadron gas at very small baryon chemical potentials, where hot medium matter is produced with almost equal numbers of particles and antiparticles. Indeed, many phenomenological models also predict that a strong first-order phase transition may occur in compressed baryon-rich matter at lower energies\cite{1,2,3,4,5,6,7,8,9}, and the purely quark-qluon plasma with zero baryonic chemical potential could be created at high energies ($\sqrt{s_{NN}}\geq13\mathrm{GeV}$)\cite{10}. Presumably, such dense baryonic matter is created at lower energies ($\sqrt{s_{NN}}<10\mathrm{GeV}$) in the central U-U collisions with beam energies $E_{lab}=0.4A\mathrm{GeV}$ ($\sqrt{s_{NN}}=2.07\mathrm{GeV}$) at the Cooling Storage Rrings (CSR) at the Heavy Ion Research Facility in Lanzhou (HIRFL)\cite{11}, the central U-U collisions with beam energies $E_{lab}=1.0A\mathrm{GeV}$ ($\sqrt{s_{NN}}=2.32\mathrm{GeV}$) at High Intensity Heavy Ion Accelerator Facility (HIAF)\cite{12}, the central Au-Au collisions ($\sqrt{s_{NN}}=2.7-4.9\mathrm{GeV}$) of Compressed Baryonic Matter (CBM) experiment at Facility for Antiproton and Ion Research (FAIR)\cite{13,14}, the central Au-Au collisions ($\sqrt{s_{NN}}<10\mathrm{GeV}$) of Multi-Purpose Detector (MPD) experiment at Nuclotron-based Ion Collider fAcility (NICA)\cite{15}, and the central Au-Au collisions ($\sqrt{s_{NN}}=7.7\mathrm{GeV}$) of the Beam Energy Scan (BES) program at Relativistic Heavy Ion Collider (RHIC)\cite{16,17}. Especially for the center-of-mass energies $\sqrt{s_{NN}}\leq5\mathrm{GeV}$, the dense baryonic matter is dominated by the incoming nucleons\cite{18,19}.

In the present work, we use the relativistic perfect-fluid hydrodynamics to describe the space-time evolution of dense baryonic matter produced in the central nucleus-nucleus collisions at HIRFL-CSR, HIAF, FAIR-CBM, NICA-MPD, and RHIC-BES. In relativistic heavy-ion collisions, the relativistic hydrodynamical equations describe the collective properties of dense baryonic matter. The Bjorken solution\cite{20} provides an estimate of the (1+1)-dimensional longitudinal expansion of the fireball in the context of the Landau hydrodynamic model\cite{21,22}, but with a different initial boundary condition. The transverse-flow effects have been studied by relativistic perfect hydrodynamics that assumes cylindrical symmetry along the transverse direction and boost invariant along the longitudinal direction\cite{23,24,25,26,27,28,29,30,31,32}, as well as described by the relativistic simple wave of Riemann irrespective of the initial conditions\cite{33,34,35}. After the initial proper time $\tau_{i}$ and initial temperature $T_{i}$, the dense baryonic matter is regarded as thermalized. The fireball temperature $T(\tau,r)$ is given as a function of proper time $\tau$ and radial distance $r$ by the calculation of the transverse-flow. Moreover, we also present the first preliminary calculation on the production of low-mass dileptons and low-$p_{T}$ photons from the pion-pion interaction in dense baryonic matter produced in the central Au-Au and U-U collisions. We find the suppression of the transverse-flow effects for the low-mass dileptons and low-$p_{T}$ photons production from the dense baryonic matter is also apparent at lower energies at HIRFL-CSR, HIAF, FAIR-CBM, NICA-MPD, and RHIC-BES.

In the case of the central nucleus-nucleus collisions at HIRFL-CSR, HIAF, FAIR-CBM, NICA-MPD, and RHIC-BES, we shall assume that the dense baryonic matter with low temperature and high baryon density produced at low energies can be treated as a perfect-fluid, and the freeze-out is performed while the energy density of the fireball drops below $\rho_{0}\approx0.16\mathrm{GeV}/\mathrm{fm}^{3}$. In the relativistic perfect-fluid hydrodynamics model the grand canonical ensemble partition function contains all relevant degrees of freedom of the dense baryonic matter and and implicitly includes interactions that result in fireball formation. The logarithm of the grand canonical ensemble partition function of baryons and mesons with mass, chemical potential, and degeneracy factor in the large volume limit of the dense baryonic matter (BM) produced in the central nucleus-nucleus collisions at HIRFL-CSR, HIAF, FAIR-CBM, NICA-MPD, and RHIC-BES can be written as\cite{36,37}
\begin{eqnarray}\label{y1}
&&\!\!\!\!\!\!\!\!T\ln Z(T,\mu,V)\bigg|_{BM}\!\!\!=\!T\ln Z(T,\mu,V)\bigg|_{N}\!\!\!+\!T\ln Z(T,\mu,V)\bigg|_{M}\nonumber\\[1mm]
&=&\frac{g_{N}V}{6\pi^{2}}\!\!\int_{0}^{\infty}\!\!\!\!\frac{dkk^{4}}{(k^{2}\!+\!M^{2})^{1/2}}\frac{1}{\exp\{[(k^{2}\!+\!M^{2})^{1/2}\!-\!\mu_{B}]/T\}\!+\!1}\nonumber\\[1mm]
&&+\frac{g_{m}V}{6\pi^{2}}\!\!\int_{0}^{\infty}\!\!\!\!\frac{dkk^{4}}{(k^{2}+m^{2})^{1/2}}\frac{1}{\exp[(k^{2}+m^{2})^{1/2}/T]-1},
\end{eqnarray}
where $g_{N}=4$ is the degeneracy factor of the baryon, and $g_{m}=3$ is the degeneracy factor of the meson.

The Fermi-Dirac distribution function of baryons and Bose-Einstein distribution function of mesons in the dense baryonic matter are given by
\begin{eqnarray}\label{y2}
&&f_{B}(E,T,\mu_{B})=\frac{1}{\exp\{[(k^{2}+M^{2})^{1/2}-\mu_{B}]/T\}+1},\nonumber\\[1mm]
&&f_{m}(E,T)=\frac{1}{\exp[(k^{2}+m^{2})^{1/2}/T]-1},
\end{eqnarray}
where $\mu_{B}$ is the energy dependence of the baryon chemical potential\cite{38,39,40,41}.

The entropy density, number density, energy density, and pressure associated with these distributions in the dense baryonic matter can be expressed as
\begin{eqnarray}\label{y3}
\!\!\!\!s&=&\frac{3g_{N}M^{2}T}{2\pi^{2}}\sum_{n=1}^{\infty}\frac{(-1)^{n-1}}{n^{2}}K_{2}\left(\frac{nM}{T}\right)\exp(n\mu_{B}/T)\nonumber\\[1mm]
&&\!\!\!\!\!\!\!\!\!\!\!+\frac{g_{N}M^{3}}{2\pi^{2}}\sum_{n=1}^{\infty}\frac{(-1)^{n-1}}{n}K_{1}\left(\frac{nM}{T}\right)\exp(n\mu_{B}/T)\nonumber\\[1mm]
&&\!\!\!\!\!\!\!\!\!\!\!-\frac{g_{N}M^{2}\mu_{B}}{2\pi^{2}}\sum_{n=1}^{\infty}\frac{(-1)^{n-1}}{n}K_{2}\left(\frac{nM}{T}\right)\exp(n\mu_{B}/T)\nonumber\\[1mm]
&&\!\!\!\!\!\!\!\!\!\!\!+\frac{g_{m}m^{2}}{2\pi^{2}}\left[3T\!\!\sum_{n=1}^{\infty}\!\!\frac{1}{n^{2}}K_{2}\left(\frac{nm}{T}\right)\!+\!m\!\!\sum_{n=1}^{\infty}\!\!\frac{1}{n}K_{1}\left(\frac{nm}{T}\right)\right],
\end{eqnarray}
\begin{eqnarray}\label{y4}
\!\!\!\!\!\!\!\!n=\frac{g_{N}M^{2}T}{2\pi^{2}}\!\sum_{n=1}^{\infty}\!\frac{(-1)^{n-1}}{n^{2}}K_{2}\!\!\left(\frac{nM}{T}\right)\exp(n\mu_{B}/T),
\end{eqnarray}
\begin{eqnarray}\label{y5}
\!\!\!\!\epsilon&=&\frac{3g_{N}M^{2}T^{2}}{2\pi^{2}}\sum_{n=1}^{\infty}\frac{(-1)^{n-1}}{n^{2}}K_{2}\left(\frac{nM}{T}\right)\exp(n\mu_{B}/T)\nonumber\\[1mm]
&&\!\!\!\!\!\!\!\!\!\!+\frac{g_{N}M^{3}T}{2\pi^{2}}\sum_{n=1}^{\infty}\frac{(-1)^{n-1}}{n}K_{1}\left(\frac{nM}{T}\right)\exp(n\mu_{B}/T)\nonumber\\[1mm]
&&\!\!\!\!\!\!\!\!\!\!+\frac{g_{m}m^{2}T}{2\pi^{2}}\!\!\left[3T\!\!\sum_{n=1}^{\infty}\!\!\frac{1}{n^{2}}K_{2}\!\left(\frac{nm}{T}\right)\!+\!m\!\!\sum_{n=1}^{\infty}\!\!\frac{1}{n}K_{1}\!\left(\frac{nm}{T}\right)\right],
\end{eqnarray}
\begin{eqnarray}\label{y6}
P&=&\frac{g_{N}M^{2}T^{2}}{2\pi^{2}}\sum_{n=1}^{\infty}\frac{(-1)^{n-1}}{n^{2}}K_{2}\left(\frac{nM}{T}\right)\exp(n\mu_{B}/T)\nonumber\\[1mm]
&&+\frac{g_{m}m^{2}T^{2}}{2\pi^{2}}\sum_{n=1}^{\infty}\frac{1}{n^{2}}K_{2}\left(\frac{nm}{T}\right),
\end{eqnarray}
where $K_{\nu}(x)$ is the modified Bessel function.

Then the perfect-fluid hydrodynamics equations of the motion for the dense baryonic matter produced in the central nucleus-nucleus collisions at HIRFL-CSR, HIAF, FAIR-CBM, NICA-MPD, and RHIC-BES energies can be written as
\begin{eqnarray}\label{y7}
\partial_{\mu}T^{\mu\nu}=0,
\end{eqnarray}
where the energy-momentum tensor is given by
\begin{eqnarray}\label{y8}
T^{\mu\nu}=(\epsilon+P)u^{\mu}u^{\nu}-g^{\mu\nu}P,
\end{eqnarray}
where $\epsilon$ is the energy density, $P$ is the pressure, and $u^{\mu}$ is the four-velocity of the collective flow,
\begin{eqnarray}\label{y9}
u^{\mu}=(\gamma,\gamma\mathbf{v}),
\end{eqnarray}
where $\gamma=(1-v^{2})^{-1/2}$, $g^{\mu\nu}=\mathrm{diag}(1,-1,-1,-1)$ is the metric tensor, and the space-time coordinate denoted by $x_{\mu}=(t,r)$. We neglect the viscous drag of fireball, and thus the expansion is isentropic, which leads to
\begin{eqnarray}\label{y10}
\partial_{\mu}s^{\mu}=0,s^{\mu}=su^{\mu},
\end{eqnarray}
\begin{eqnarray}\label{y11}
\partial_{\mu}J^{\mu}=0,J^{\mu}=nu^{\mu},
\end{eqnarray}
where $s$ and $n$ are the entropy density and number density of the fireball, respectively.

For the (1+1)-dimensional Bjorken longitudinal expansion of the dense baryonic matter, the dynamics evolution equation for the fireball is given by
\begin{eqnarray}\label{y12}
\frac{\partial\epsilon}{\partial\tau}+\frac{\epsilon+P}{\tau}=0,
\end{eqnarray}
then the time evolution for the temperature of the dense baryonic matter can be written as
\begin{eqnarray}\label{y13}
T(\tau)=T(\tau_{i})\left(\frac{M-\mu_{B}(\tau)}{M-\mu_{B}(\tau_{i})}\right)^{2}\left(\frac{\tau_{i}}{\tau}\right)^{(1+c_{s}^{2})/3},
\end{eqnarray}
if the baryonic chemical potential is a constant, $\mu_{B}(\tau)\approx\mu_{B}(\tau_{i})\approx\mathrm{Constant}$, then
\begin{eqnarray}\label{y14}
T(\tau)=T(\tau_{i})\left(\frac{\tau_{i}}{\tau}\right)^{(1+c_{s}^{2})/3},
\end{eqnarray}
where $M$ is the mass of nucleon, and $c_{s}^{2}=0.05-0.15$ is the speed of sound.

\begin{table*}
\centering\caption{The initial conditions of the perfect-fluid hydrodynamical expansion for the dense baryonic matter produced at low energy central nucleus-nucleus collisions ($\sqrt{s_{NN}}<10\mathrm{GeV}$) at HIRFL-CSR, HIAF, FAIR-CBM, NICA-MPD, and RHIC-BES.}
\label{T1}
\begin{tabular}{ c|c c c c c }
  \hline
  \quad\quad center-of-mass energy\quad\quad\quad & \quad$\mu_{B}(\mathrm{MeV})$\quad\quad & $T_{i}(\mathrm{MeV})$\quad & \quad$\tau_{i}(\mathrm{fm/c})$\quad & \quad$\epsilon_{i}(\mathrm{GeV/fm}^{3})$\quad & \quad dN/dy \\
  \hline
  HIRFL-CSR,$\sqrt{s_{NN}}=2.07\mathrm{GeV}$ & 836 & 43.2 & \quad1.2 & \quad0.3383 & \quad400 \\
  \quad\quad\quad HIAF,$\sqrt{s_{NN}}=2.32\mathrm{GeV}$ & 800 & 54.9 & \quad1.2 & \quad0.3797 & \quad291 \\
  \!FAIR-CBM,$\sqrt{s_{NN}}=2.7\mathrm{GeV}$ & 740 & 54.5 & \quad1.2 & \quad0.3704 & \quad282 \\
  \!FAIR-CBM,$\sqrt{s_{NN}}=4.9\mathrm{GeV}$ & 548 & 143.6 & \quad0.46 & \quad0.6722 & \quad34 \\
  \!\!\!\,NICA-MPD,$\sqrt{s_{NN}}=4.0\mathrm{GeV}$ & 625 & 107.2 & \quad0.613 & \quad0.5487 & \quad55 \\
  \!\!\!\,NICA-MPD,$\sqrt{s_{NN}}=7.0\mathrm{GeV}$ & 450 & 149.4 & \quad0.44 & \quad0.9603 & \quad23 \\
  RHIC-BES,$\sqrt{s_{NN}}=7.7\mathrm{GeV}$ & 422 & 186.2 & \quad0.35 & \quad1.0563 & \quad21 \\
  \hline
\end{tabular}
\end{table*}

Moreover, in the perfect-fluid with cylindrical symmetry along the transverse direction and boost invariant along the longitudinal direction together with the fluid velocity vector satisfying the constraint $u_{\mu}u^{\mu}=1$, requires that the fluid four-velocity for cylindrical symmetry be of the form\cite{24}
\begin{eqnarray}\label{y15}
u^{\mu}=\gamma_{r}(\tau,r)(t/\tau,v_{r}(\tau,r),z/\tau),
\end{eqnarray}
where we have used
\begin{eqnarray}\label{y16}
\gamma_{r}(\tau,r)=[1-v_{r}^{2}(\tau,r)]^{-1/2},
\end{eqnarray}
\begin{eqnarray}\label{y17}
\tau=(t^{2}-z^{2})^{1/2},
\end{eqnarray}
\begin{eqnarray}\label{y18}
y=\frac{1}{2}\ln\frac{t+z}{t-z},
\end{eqnarray}
then Lorentz-scalar quantities such as the entropy density, number density, energy density, pressure, and radial velocity $v_{r}(\tau,r)$ are functions only of $\tau$ and $r$, with no dependence upon the space-time rapidity $y$.

\begin{figure}[b]
\includegraphics[width=9cm,height=7cm]{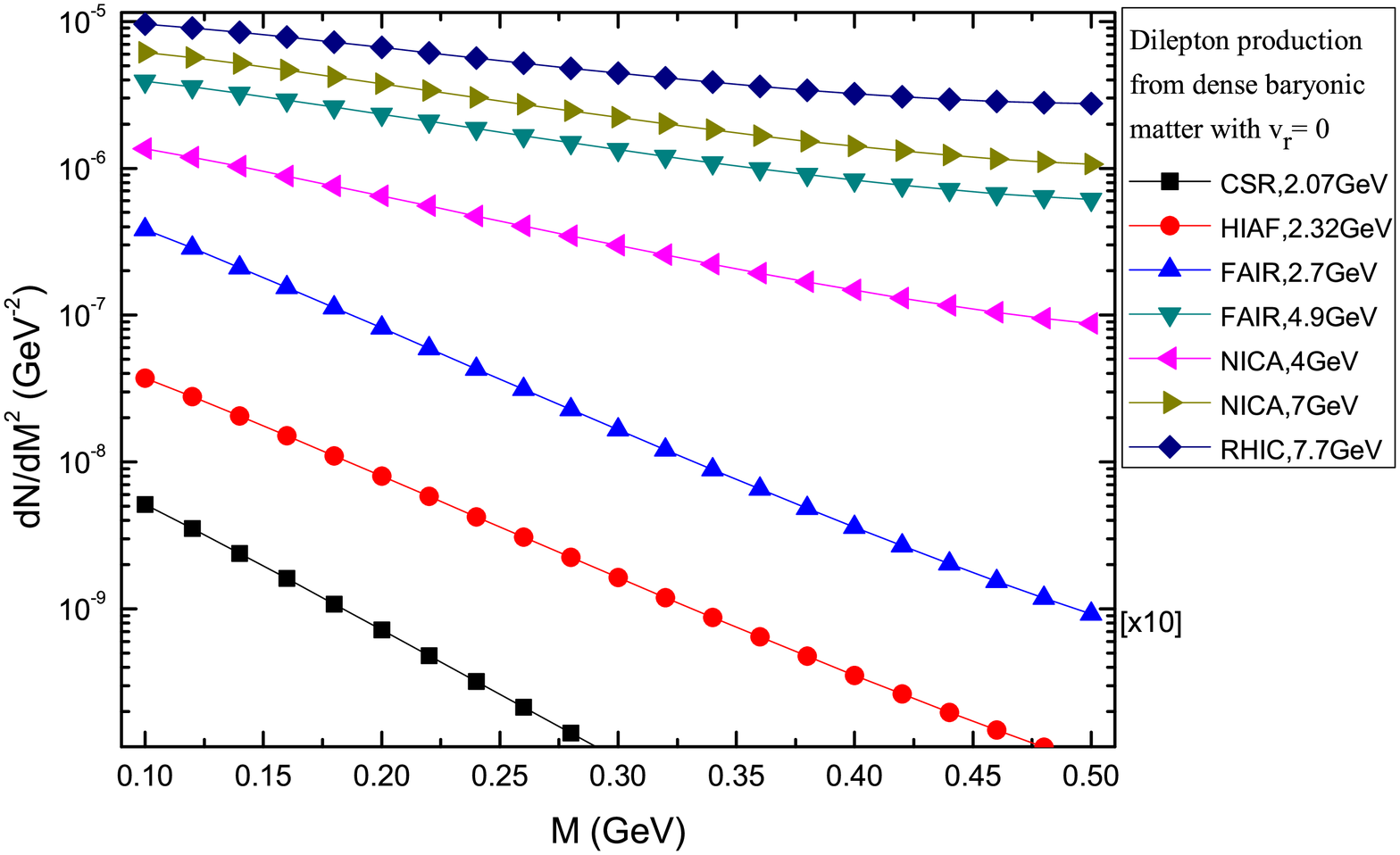}
\caption{\label{Fig1}  The low-maa dileptons yield for the radial velocity $v_{r}(\tau,r)=0$ in the dense baryonic matter at HIRFL-CSR, HIAF, FAIR-CBM, NICA-MPD, and RHIC-BES energies.}
\end{figure}
\begin{figure}[b]
\includegraphics[width=9cm,height=7cm]{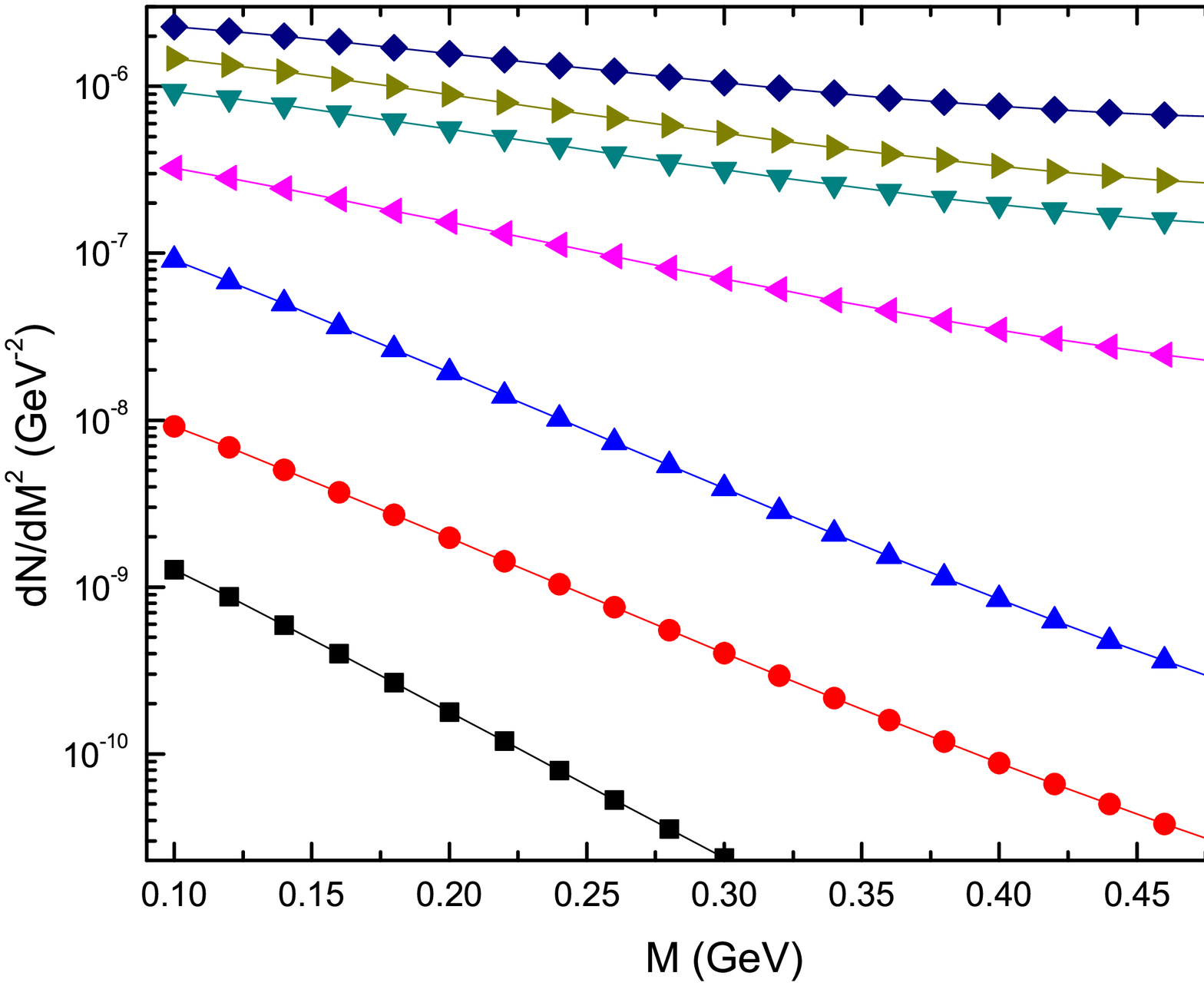}
\caption{\label{Fig2}  The low-mass dileptons yield for the radial velocity $v_{r}(\tau,r)\neq0$ in the dense baryonic matter at HIRFL-CSR, HIAF, FAIR-CBM, NICA-MPD, and RHIC-BES energies.}
\end{figure}

Then the perfect-fluid hydrodynamic equations of motion of the dense baryonic matter for the (3+1)-dimensional expansion with cylindrical symmetry along the transverse direction and boost invariance along the longitudinal direction can be expressed as
\begin{eqnarray}\label{y19}
&&\!\!\!\!\!\!\!\!\frac{\partial\epsilon}{\partial\tau}+\frac{(\epsilon+P)}{\tau}+(\epsilon+P)\frac{\partial v_{r}(\tau,r)}{\partial r}\nonumber\\[1mm]
&&\!\!\!\!\!\!\!\!+(\epsilon+P)\left[v_{r}(\tau,r)\frac{\partial\ln\gamma_{r}(\tau,r)}{\partial r}+\frac{\partial\ln\gamma_{r}(\tau,r)}{\partial\tau}\right]=0,
\end{eqnarray}
where the radial velocity is given by\cite{42,43,44,45,46}
\begin{eqnarray}\label{y20}
v_{r}(\tau,r)=\frac{2q^{2}r\tau}{1+(q\tau)^{2}+(qr)^{2}},
\end{eqnarray}
where the free parameter $q$ with energy dimensions is allowed one to specify a scale which is related to the transverse size of the fireball $q=R^{-1}$. In the case of $v_{r}(\tau,r)=0$, the (1+1)-dimensional Bjorken equation of the dense baryonic matter is recovered.

Then the time evolution for the temperature of the dense baryonic matter with the transverse-flow effects can be written as
\begin{eqnarray}\label{y21}
T(\tau,r)\!=\!T(\tau_{i},r)\!\!\left[\!\frac{M\!\!-\!\!\mu_{B}(\tau)}{M\!\!-\!\!\mu_{B}(\tau_{i})}\!\right]^{\!2}\!\!\left[\!\frac{\tau_{i}(q^{-2}\!+\!r^{2}\!+\!\tau_{i}^{2})}{\tau(q^{-2}\!+\!r^{2}\!+\!\tau^{2})}\!\right]^{\!\!(1+c_{s}^{2})/3},\nonumber\\[1mm]
\end{eqnarray}
if the baryonic chemical potential is a constant, $\mu_{B}(\tau)\approx\mu_{B}(\tau_{i})\approx\mathrm{Constant}$, then
\begin{eqnarray}\label{y22}
T(\tau,r)=T(\tau_{i},r)\left[\frac{\tau_{i}(q^{-2}+r^{2}+\tau_{i}^{2})}{\tau(q^{-2}+r^{2}+\tau^{2})}\right]^{(1+c_{s}^{2})/3},
\end{eqnarray}
where $M$ is the mass of nucleon, and $c_{s}^{2}$ is the speed of sound.

\begin{figure}[t]
\includegraphics[width=9cm,height=7cm]{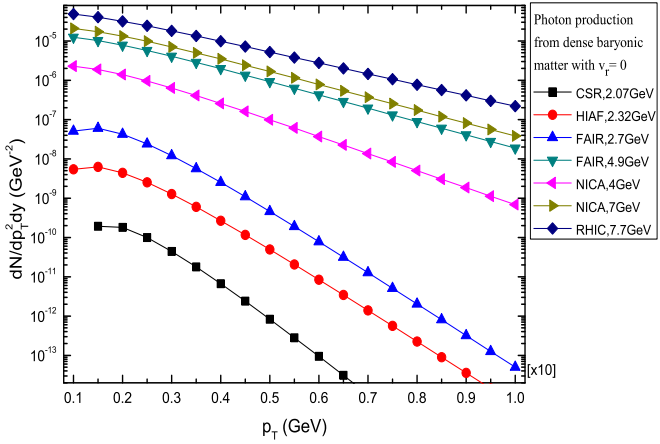}
\caption{\label{Fig3}  The low-$p_{T}$ photons yield for the radial velocity $v_{r}(\tau,r)=0$ in the dense baryonic matter at HIRFL-CSR, HIAF, FAIR-CBM, NICA-MPD, and RHIC-BES energies.}
\end{figure}

The initial conditions of the perfect-fluid hydrodynamical expansion for the dense baryonic matter produced at low energy central nucleus-nucleus collisions ($\sqrt{s_{NN}}<10\mathrm{GeV}$) at HIRFL-CSR, HIAF, FAIR-CBM, NICA-MPD, and RHIC-BES are presented in Table \ref{T1}. The initial conditions of the transverse expansion for (3+1)-dimensional perfect-fluid hydrodynamics are also chosen such that the radial velocity $v_{r}(\tau_{i},r)=0$ along with a given initial temperature $T(\tau_{i},r)=T_{i}$ within the transverse radius\cite{29}.

\begin{figure}[t]
\includegraphics[width=9cm,height=7cm]{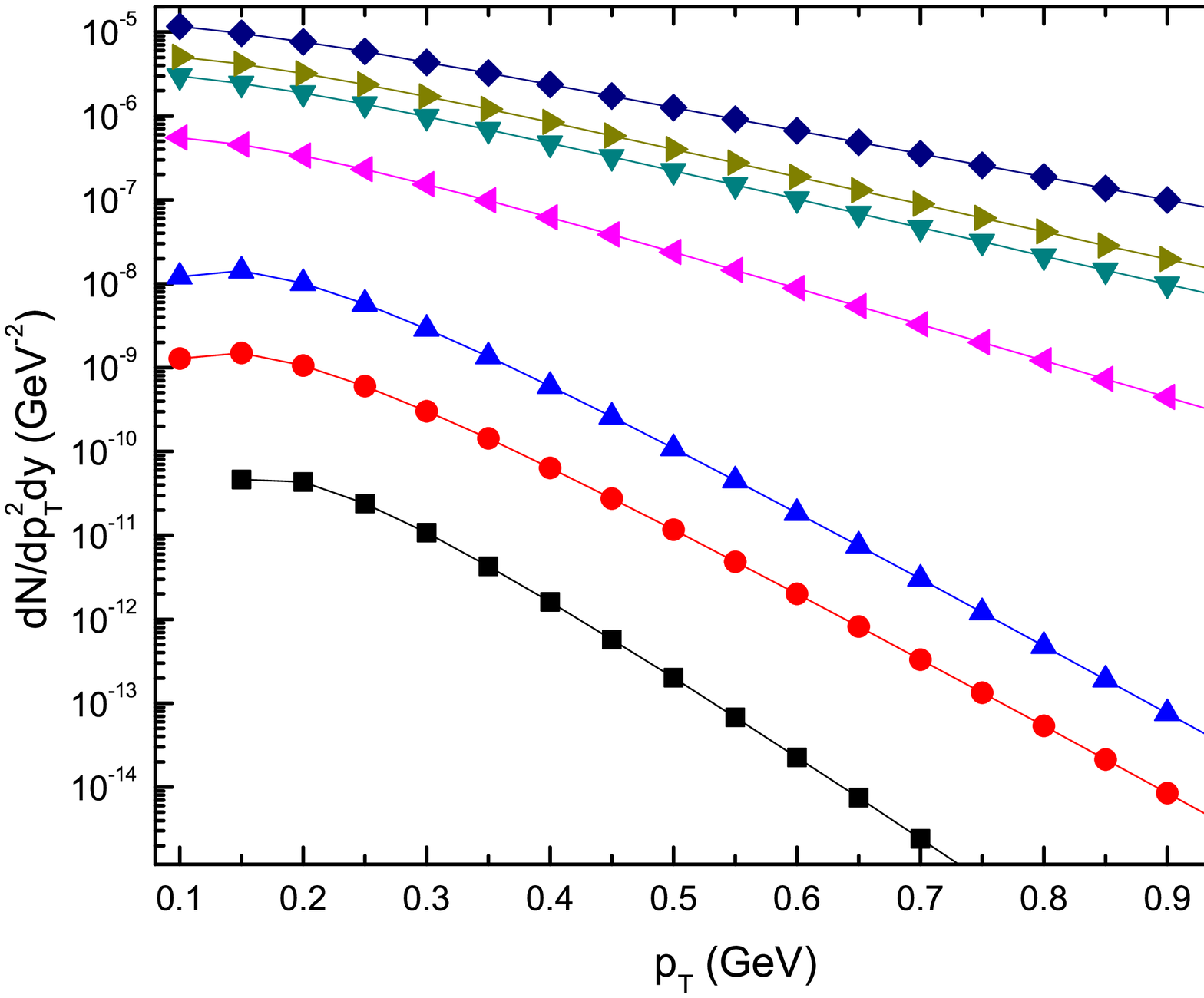}
\caption{\label{Fig4}  The low-$p_{T}$ photons yield for the radial velocity $v_{r}(\tau,r)\neq0$ in the dense baryonic matter at HIRFL-CSR, HIAF, FAIR-CBM, NICA-MPD, and RHIC-BES energies.}
\end{figure}

Moreover, we also present the first preliminary calculation on the production of low-mass dileptons and low-$p_{T}$ photons from the pion-pion interaction in dense baryonic matter produced in the central Au-Au and U-U collisions. In the central nucleus-nucleus collisions, the yield of low-mass dileptons produced by $\pi\pi\rightarrow V\rightarrow\ell^{+}\ell^{-}$ processes in the dense baryonic matter can be written as
\begin{eqnarray}\label{y23}
\!\!\!\!\!\!\!\!\!\!\!\!\frac{dN_{\pi\pi\rightarrow\ell^{+}\ell^{-}}}{dM^{2}d^{4}x}\!=\!\frac{\hat{\sigma}(M)}{2(2\pi)^{4}}M^{2}\!\!\left(1-\frac{4M_{\pi}^{2}}{M^{2}}\right)^{\!\!1/2}\!\!\!TMK_{1}\!\!\left(\frac{M}{T}\right)\!\!,
\end{eqnarray}
where the cross section of $\pi\pi\rightarrow V\rightarrow\ell^{+}\ell^{-}$ processes is given by
\begin{eqnarray}\label{y24}
\hat{\sigma}(M)\!=\!\frac{4\pi}{3}\frac{\alpha^{2}}{M^{2}}\!\!\left(\!\!1\!-\!\frac{4m_{\pi}^{2}}{M^{2}}\!\right)^{\!\!\!1\!/\!2}\!\!\!\!\left(\!\!1\!-\!\frac{4m_{\ell}^{2}}{M^{2}}\!\right)^{\!\!\!1\!/\!2}\!\!\!\!\left(\!\!1\!+\!\frac{2m_{\ell}^{2}}{M^{2}}\!\right)^{\!\!\!1\!/\!2}\!\!\!\!\!\left|\!F_{\pi}(m_{V})\!\right|^{2}\!,\nonumber\\[1mm]
\end{eqnarray}
where the form factor can be expressed as\cite{47,48}
\begin{eqnarray}\label{y25}
\left|F_{\pi}(m_{V})\right|^{2}=\frac{m_{V}^{4}}{(M^{2}-m_{V}^{2})^{2}-\Gamma_{V}m_{V}^{2}},
\end{eqnarray}
where $M$ is the invariant mass of the dileptons. Indeed, the yield of the low-$p_{T}$ photons produced by $\pi\pi\rightarrow\gamma\gamma$ processes in dense baryonic matter can be written as
\begin{eqnarray}\label{y25}
\!\!\!\!\!\!\!\!\!\!E\frac{dN}{d^{3}pd^{4}x}(\pi\pi\!\rightarrow\!\gamma\gamma)=\frac{\pi^{3}\alpha^{2}}{(2\pi)^{5}}f_{\pi}(\mathbf{p})\left[\ln\frac{ET}{m_{\pi}^{2}}+0.938\right],
\end{eqnarray}
where $f_{\pi}(\mathbf{p})$ is the Bose-Einstein distribution function of pions in the dense baryonic matter.

In our analysis, we assume that the space-time evolution of dense baryonic matter produced in the low energy central nucleus-nucleus collisions at HIRFL-CSR, HIAF, FAIR-CBM, NICA-MPD, and RHIC-BES can be described by the relativistic perfect-fluid hydrodynamic. If there is only a longitudinal expansion of dense baryonic matter, it is the well-known Bjorken expansion. However the transverse-flow effects with cylindrical symmetry along the transverse direction and boost invariant along the longitudinal direction can also not be negligible. The reasonable values of the initial conditions such as the baryonic chemical potential, initial time, initial temperature, initial energy density, and rapidity distribution as shown in Table \ref{T1} can be obtained. we also present the first preliminary calculation on the production of low-mass dileptons and low-$p_{T}$ photons that can be considered as the signals of the dense baryonic matter for the case of radial velocity $v_{r}(\tau_{i},r)=0$ are plotted in Fig. \ref{Fig1} and Fig. \ref{Fig3}, and the case of radial velocity $v_{r}(\tau_{i},r)\neq0$ are plotted in Fig. \ref{Fig2} and Fig. \ref{Fig4}. We find that the suppression of the low-mass dileptons and low-$p_{T}$ photons production from the transverse flow effect can not be negligible.

In summary, we have studied the space-time evolution of dense baryonic matter by the relativistic perfect-fluid hydrodynamics in the low energy central nucleus-nucleus collisions at HIRFL-CSR, HIAF, FAIR-CBM, NICA-MPD, and RHIC-BES. The transverse-flow effects of the fireball with cylindrical symmetry along the transverse direction and boost invariant along the longitudinal direction is also considered. We also present the first preliminary calculation on the low-mass dileptons and low-$p_{T}$ photons yield that is suppressed by the transverse-flow effects (3+1)-dimensional hydrodynamical expansion for the dense baryonic matter at HIRFL-CSR, HIAF, FAIR-CBM, NICA-MPD, and RHIC-BES energies.

This work is supported by the National Basic Research Program of China (973 Program) with Grant No. 2014CB845405 and the China Postdoctoral Science Foundation funded project with Grant No. 2017M610663.







\bibliographystyle{model1-num-names}
\bibliography{<your-bib-database>}

\begin{thebibliography}{99}


\bibitem{1}
M. G. Alford, K. Rajagopal, and F. Wilzcek, Phys. Lett. B {422}, {247} (1998).

\bibitem{2}
O. Scavenius, A. Mocsy, I. N. Mishustin, and D. H. Rischke, Phys. Rev. C {64}, {045202} (2001).

\bibitem{3}
M. A. Stephanov, Prog. Theor. Phys. Suppl. {153}, {139} (2004).

\bibitem{4}
M. A. Stephanov, Int. J. Mod. Phys. A {20}, {4387} (2005).

\bibitem{5}
A. V. Merdeev, L. M. Satarov, and I. N. Mishustin, Phys. Rev. C {84}, {014907} (2011).

\bibitem{6}
M. Lorenz et al. (HADES Collaboration), Nucl. Phys. A {931}, {785} (2014).

\bibitem{7}
P. Batyuk, D. Blaschke, M. Bleicher, Yu. B. Ivanov, Iu. Karpenko, S. Merts, M. Nahrgang, H. Petersen, and O. Rogachevsky, Phys. Rev. C {94}, {044917} (2016).

\bibitem{8}
V. Vovchenko, V. V. Begun, and M. I. Gorenstein, Phys. Rev. C {93}, {064906} (2016)

\bibitem{9}
T. Ablyazimov et al. (CBM Collaboration), Eur. Phys. J. A {53}, {60} (2017).

\bibitem{10}
K. Kajantie, J. Kapusta, L. McLerran, and A. Mekjian, Phys. Rev. D {34}, {2746} (1986).

\bibitem{11}
Y. Y. Wang, D. T. Zhou, J. F. Luo, J. C. Zhang, W. X. Zhou, F. F. Ni, J. Yin, J. Yin, Y. J. Yuan, and J. B. Shang-Guan, Chin. Phys. C {40}, {047003} (2016).

\bibitem{12}
G. D. Shen, J. C. Yang, J. W. Xia, L. J. Mao, D. Y. Yin, W. P. Chai, J. Shi, L. N. Sheng, A. Smirnov, B. Wu, and H. Zhao, Chin. Phys. C {40}, {087004} (2016).

\bibitem{13}
B. Friman, C. Hohne, J. Knoll, S. Leupold, J. Randrup, R. Rapp, and P. Senger, Lect. Notes Phys. {814}, {1} (2011).

\bibitem{14}
P. Senger et al. (CBM Collaboration), Cent. Eur. J. Phys. {10}, {1289} (2012).

\bibitem{15}
NICA White Paper [http://theor0.jinr.ru/twiki-cgi/view/NICA/WebHome].

\bibitem{16}
S. Das et al. (STAR collaboration), EPJ Web Conf. {90}, {08007} (2015).

\bibitem{17}
J. T. Mitchell et al. (PHENIX Collaboration), Nucl. Phys. A {956}, {842} (2016).

\bibitem{18}
A. Andronic, P. Braun-Munzinger, and J. Stachel, Nucl. Phys. A {772}, {167} (2006).

\bibitem{19}
H. Stoecker, I. Augustin, J. Steinheimer, A. Andronic, T. Saito, P. Senger, Nucl. Phys. A {827}, {624c} (2009).

\bibitem{20}
J. D. Bjorken, Phys. Rev. D {27}, {140} (1983).

\bibitem{21}
L. D. Landau, Izv. Akad. Nauk SSSR {17}, {51} (1953).

\bibitem{22}
S. Z. Belenkii, L. D. Landau, Usp. Fiz. Nauk {56}, {309} (1955).

\bibitem{23}
R. C. Hwa and K. Kajantie, Phys. Rev. D {32}, {1109} (1985).

\bibitem{24}
H. von Gersdorff, L.McLerran, M. Kataja, and P. V. Ruuskanen, Phys. Rev. D {34}, {794} (1986).

\bibitem{25}
K. Kajantie,M. Kataja, L. McLerran, and P.V. Ruuskanen, Phys. Rev. D {34}, {811} (1986).

\bibitem{26}
B. L. Friman, K. Kajantie, and P.V. Ruuskanen, Nucl. Phys. B {266}, {468} (1986).

\bibitem{27}
K. S. Lee, U. Heinz, E. Schnedermann, Z. Phys. C {48}, {525} (1990).

\bibitem{28}
P. L$\acute{\mathrm{e}}$vai and B. M$\ddot{\mathrm{u}}$ller, Phys. Rev. Lett. {67}, {1519} (1991).

\bibitem{29}
Jan-e Alam, D. K. Srivastava, B. Sinha, and D. N. Basu, Phys. Rev. D {48}, {1117} (1993).

\bibitem{30}
S. Esumi, S. Chapman, H. van Hecke, and N. Xu, Phys. Rev. C {55}, {R2163} (1997).

\bibitem{31}
R. Ryblewski, Acta Phys. Pol. B {40}, {2019} (2009).

\bibitem{32}
Y. P. Fu and Q. Xi, Phys. Rev. C {92}, {024914} (2015).

\bibitem{33}
G. Baym, B. L. Friman, J.-P. Blaizot, M. Soyeur, and W. Czyz, Nucl. Phys. A {407}, {541} (1983).

\bibitem{34}
A. Bialas, W. Czyz, and A. Kolawa, Acta Phys. Pol. A {15}, {229} (1984).

\bibitem{35}
A. Bialas and W. Czyz, Acta Phys. Pol. A {15}, {247} (1984).

\bibitem{36}
J. Cleymans, R.V. Gavai, E. Suhonen, Phys. Rept. {130}, {217} (1986).

\bibitem{37}
R. Vogt, \emph{Ultrarelativistic heavy-ion collisions} (Oxford: Elsevier Besloten Vennootschap, 2007).

\bibitem{38}
J. Cleymans, H. Oeschler, K. Redlich, and S. Wheaton, Phys. Rev. C {73}, {034905} (2006).

\bibitem{39}
A. Kadeer, J. G. Korner, and U. Moosbrugger, Eur. Phys. J. C {59}, {27} (2009).

\bibitem{40}
F. Karsch and K. Redlich, Phys. Lett. B {695}, {136} (2011).

\bibitem{41}
P. Garg, D. K. Mishra, P. K. Netrakanti, B. Mohanty, A. K. Mohanty, B. K. Singh, and N. Xu, Phys. Lett. B {726}, {691} (2013).

\bibitem{42}
S. S. Gubser, Phys. Rev. D {82}, {085027} (2010).

\bibitem{43}
S. S. Gubser and A. Yarom, Nucl. Phys. B {846}, {469} (2011).

\bibitem{44}
P. Staig and E. Shuryak, Phys. Rev. C {84}, {044912} (2011).

\bibitem{45}
J. Kapusta, B. Muller, and M. Stephanov, Phys. Rev. C {85}, {054906} (2012).

\bibitem{46}
T. Springer and M. Stephanov, Nucl. Phys. A {904-905}, {1027c} (2013).

\bibitem{47}
C. Y. Wong, \emph{Introduction to High Energy Heavy-Ion Collisions} (World Scientific, 1994).

\bibitem{48}
G. J. Gounaris and J. J. Sakurai, Phys. Rev. Lett. {21}, {244} (1968).
















\end{thebibliography}

\end{document}